\documentstyle[prl,aps,twocolumn,epsf]{revtex}

\newcommand{\be}{\begin{equation}}
\newcommand{\ee}{\end{equation}}
\newcommand{\bea}{\begin{eqnarray}}
\newcommand{\eea}{\end{eqnarray}}

\newcommand{\pd}{{\partial}}
\newcommand{\phizc}{\phi_{0c}}
\newcommand{\phipc}{\phi_{+c}}
\newcommand{\phim}{\phi_{\rm max}}
\newcommand{\Tm}{\Theta_{\rm max}}

\begin{document}

\twocolumn[

\hsize\textwidth\columnwidth\hsize\csname @twocolumnfalse\endcsname

\draft

\title{Weakly First Order Cosmological Phase Transitions and Fermion
Production}


\author{Marcelo Gleiser[1] and Mark Trodden[2]}

\address{[1] Department of Physics and Astronomy, Dartmouth College,
Hanover, NH 03755 USA.}

\address{[2] Department of Physics, Case Western Reserve University,
10900 Euclid Ave., Cleveland OH 44106-7079 USA.}

\date{\today}

\maketitle

\begin{abstract}
We study weakly first order cosmological phase transitions in finite
temperature field
theories. Focusing on the standard electroweak theory and its minimal
supersymmetric extension, we identify the regimes of Higgs masses for which
the phase transition in these models proceeds by significant phase mixing and
the coarsening of the subsequent domain network. This dynamics is distinct from
that for strongly first order transitions, which proceed by the nucleation
and propagation of critical bubbles. We describe how electroweak baryogenesis
might take place in these models, explaining how our new picture can relax the
sphaleron washout bound of traditional scenarios.
\end{abstract}


]

\narrowtext
Finite temperature phase transitions are of great interest in cosmology
because they provide a mechanism by which remnants of the early universe
can make an observable imprint on today's universe. Well-known examples of
this are cosmological inflation, the production of topological defects,
and electroweak baryogenesis \cite{KT}.

Cosmological phase transitions are well-understood in two particular limits.
Strongly first order phase transitions occur when a system begins in
a metastable state that is separated from a global minimum by a sufficiently
large energy barrier. In such a system, widely separated points in
space undergo quantum tunneling or thermal hopping
events, in which bubbles of the true vacuum
nucleate in the background sea of false vacuum: bubbles which
are sufficiently large that their volume energy dominates over their
surface tension expand and eventually coalesce, completing the phase
transition.
If the phase transition is second order (continuous),
the system begins in an
unstable state, with no energy barrier separation from the global minimum.
Small-amplitude, large-wavelength fluctuations grow,
followed by domain coarsening and phase separation.
This is referred to as spinodal decomposition \cite{GOLDEN}.

Away from these two extreme limits, the dynamics of phase transitions is
much less well understood. Numerical simulations
provide accurate
estimates of the strength of the phase transition, and determine the point at
which first order transitions become second order as parameters of the
theory are varied, as recently done for the standard model (SM)
\cite{kajantie}.
However, for weakly first order phase transitions an
understanding of {\it how} the phase transition proceeds and completes is
far more vague.

It is widely believed that, in the absence of exotic means for departing from
equilibrium, a strongly first order phase transition is required for
electroweak baryogenesis to occur.(For a recent review see \cite{review}.)
This is for two main reasons. If bubble
nucleation is the mechanism, then the sharp change in order parameter across
the bubble walls provides a large departure from equilibrium at each point
in space swept out by the walls. Further, the associated large energy
barrier between phases leads to thermal anomalous fermion number
violation being sufficiently suppressed in the true vacuum so that any
baryon number produced is not washed out \cite{krs}.
For a weakly first order transition
this is not the case, and the traditional criterion that the transition
be sufficiently strong provides a constraint on the theory. This constraint
can be expressed as
$\langle\phi(T_c)\rangle/T_c \geq 1$,
where $\phi$ is the Higgs field, and $T_c$ is the critical temperature for
the electroweak phase transition. This translates directly into a bound on
the mass of the Higgs boson, which, for the
minimal supersymmetric standard model (MSSM) is
$m_H^{\rm(MSSM)} < 105~{\rm GeV}$
\cite{CQW}.

In this letter, we investigate an alternative scenario for fermion
production in weakly first order phase transitions.
When transitions are weak, large-amplitude
subcritical fluctuations between the
symmetric and broken-symmetric phases cause the transition
to begin by significant phase mixing and, after inter-phase fluctuations
cease, to proceed by coarsening of the
subsequent domain network. The crucial point is that the sphaleron washout
temperature need not be the critical temperature, but the lower temperature
when
the large-amplitude fluctuations freeze out, the {\it Ginzburg temperature},
$T_G$.
Thus, fermion production and preservation may still be efficient in this
picture,
and possibly remain sufficient to account for the baryon asymmetry of the
universe. Let us describe how this might happen using the SM and
the more promising MSSM as examples.

\begin{figure}[tbp]
\caption{\label{smwashout}
Standard Model Washout.
}
\epsfxsize = 0.85 \hsize \epsfbox{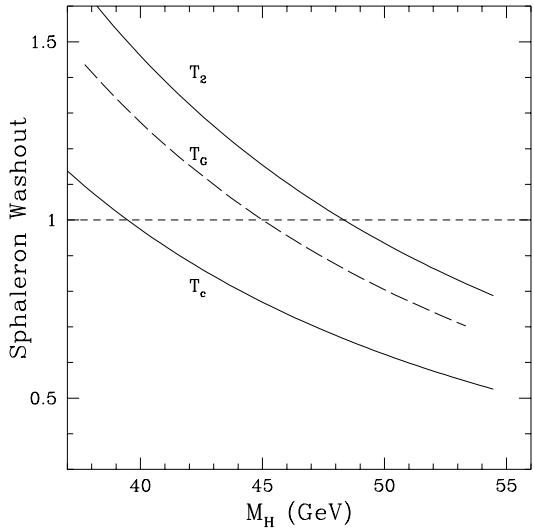}
\end{figure}

The finite temperature effective potential for the magnitude of the Higgs field
in electroweak models can often be written as
\be
V(\phi,T)=D(T^2-T_2^2)\phi^2 - ET\phi^3 +\frac{\lambda(T)}{4}\phi^4 \ ,
\label{potential}
\ee
with $T_2^2 = (m_H^2-8Bv^2)/4D$, where, for the SM,
\bea
D_{\rm SM} & = & \frac{2m_W^2+m_Z^2+2m_t^2}{8v^2} \ , \nonumber \\
E_{\rm SM} & = & \frac{2m_W^3+m_Z^3}{4\pi v^3} \ , \nonumber \\
B_{\rm SM} & = & \frac{3}{64\pi^2 v^4}(2m_W^4+m_z^4-4m_t^4) \ ,
\eea
and $\lambda(T)$ is the temperature-corrected Higgs coupling. It is meaningful
to speak of there being a global minimum, separated from a metastable minimum
by an energy barrier, for
$T_2^2<T^2<T_c^2\equiv\frac{\lambda D}{\lambda D-E^2}T_2^2$. Let us begin
by examining how the sphaleron washout temperature varies in this range, for
changing Higgs mass. In figure (\ref{smwashout}) we plot
$\langle\phi(T)\rangle/T$ as a function of the Higgs mass $m_H$, for three
values of the temperature, $T_c$, $T_G$, and $T_2$.
The lower curve on this figure demonstrates how the the traditional sphaleron
bound arises, since this curve crosses unity at a Higgs mass
around $40$ GeV. However, this, along with the upper curve shows that,
for Higgs masses in the range $40~ {\rm GeV} < m_H^{\rm SM} < 48~ {\rm GeV}$,
the temperature below which sphaleron washout is avoided may be reached
before the barrier between phases has vanished.
This is the relevant regime for our scenario.

\begin{figure}[tbp]
\caption{\label{mssmwashout}
Minimal Supersymmetric Standard Model Washout.
}
\epsfxsize = 0.85 \hsize \epsfbox{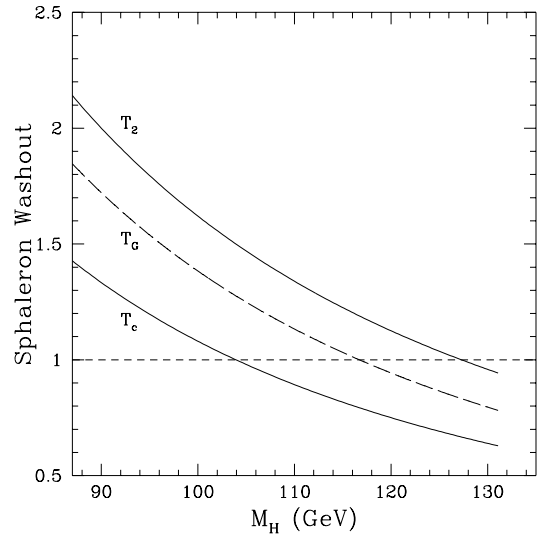}
\end{figure}

Although the SM is useful for illustrating this
effect, experimental bounds place the physical Higgs mass at $m_H>100$ GeV.
Thus, our analysis is academic in this model. However, in the MSSM, we
shall see that useful results are obtained.
For calculational convenience, in this letter we study the MSSM in the limit
in which the extra Higgs fields and all the sparticles are decoupled from
the spectrum. Thus, all that remains is the lightest Higgs and the
right-handed stop. In this limit, the one-loop finite-temperature effective
potential can be written in the same form as the SM
(\ref{potential}), but with the following parameters
\bea
D_{\rm MSSM} & = & \frac{2m_W^2+m_Z^2+4m_t^2}{8v^2} \ , \nonumber \\
E_{\rm MSSM} & = & \frac{2m_W^3+m_Z^3+2m_t^3}{4\pi v^3} \ , \nonumber \\
B_{\rm MSSM} & = & \frac{3}{64\pi^2 v^4}(2m_W^4+m_z^4-2m_t^4) \ .
\eea For this potential, we can again examine how the sphaleron
washout temperature varies with Higgs mass. We note that two-loop
corrections to the MSSM potential have been extensively discussed
in the literature. In fact, the consensus is that the inclusion of
two-loop effects enhances the strength of the transition, further
opening the window of allowed Higgs masses for efficient
baryogenesis \cite{Cline}. As our present purpose is to illustrate
a new possible mechanism for baryogenesis within weak first-order
phase transitions, we will for simplicity restrict ourselves to
one-loop corrections here, since there will always be a range of
Higgs masses for which the transition will be weakly first order.
However, we are presently working to extending our analysis to
include two-loop corrections. We plot our results for the one-loop
MSSM potential in figure (\ref{mssmwashout}).

Notice that, for the MSSM, there is a range of masses
for the lightest Higgs, within which the phase transition is
weakly first order, but with the possibility for sphaleron transitions
to become suppressed in the broken phase before the end of the transition.

The phase transition begins at the critical temperature, $T_c$. At this
temperature, when the phase transition is sufficiently weakly first order
(the sphaleron bound is violated at $T_c$ but not at $T_G$),
there are significant subcritical fluctuations, both of the broken-symmetric
(the true vacuum, $+$) and the symmetric one (false vacuum, $0$).
Consequently, significant phase mixing occurs and homogeneous nucleation
breaks down \cite{G 94}, \cite{B&G 95}.
We write the rates of the relevant subcritical
fluctuations as $G_{0\rightarrow +}$ and $G_{+\rightarrow 0}$, respectively.
As the phase-mixed plasma cools
due to the expansion of the universe, it reaches the temperature $T_G$,
at which fluctuations back to the false vacuum freeze out.
The criterion for this to occur is
\be
\left.\frac{G_{+\rightarrow 0}}{H}\right|_{T_G} =1 \ ,
\label{ginzburg}
\ee
where $H$ is the Hubble parameter. The question of how the phase transition
proceeds after this depends on how the fraction, $\gamma_+(T_G)$, of the total
volume in the broken phase at the Ginzburg temperature compares with
the percolation threshold, $p_{\rm perc}\simeq 0.31$, in three dimensions.
If $\gamma_+(T_G) < p_{\rm perc}$, we refer to {\it partial phase mixing}
initially. In this case, isolated domains of the $+$ phase will grow due to
pressure difference. However, if $\gamma_+(T_G) > p_{\rm perc}$, then we have
{\it total phase mixing} initially, and both phases percolate, being
separated by a convoluted wall, plus small isolated domains of each phase.

{}For both scenarios, baryon-number production will
depend on the ratio $\langle\phi(T_G)\rangle/T_G$, which is shown in figures
(1) and (2) as the (central) dashed line. Clearly, the details of how
baryon number will percolate into the broken-symmetric phase is determined
by how $\gamma_+(T_G)$ compares to $p_{\rm perc}$, that is, on whether
the broken-symmetric phase has percolated or not. However, our focus here is
on what we could call the necessary condition for baryogenesis to occur, which
is fixed by the ratio
$\langle\phi(T_G)\rangle/T_G$, irrespective of the details of
the phase transition. There are two possibilities, described as follows:
If this ratio is larger than one,
baryon-number violation is suppressed in the broken phase at freeze-out ($T_G$)
and a net baryonic
excess will be generated as domains of the broken phase advance on the
symmetric phase. If this ratio is smaller than one, any baryon excess will
be erased in the broken phase until the temperature drops so that the ratio
$\langle\phi(T)\rangle/T$ reaches unity, say at a temperature ${\bar T}$.
Thus, in this second case we
may expect a severe (if not deadly)
volume-suppression factor,
unless the domains advance sufficiently slowly
so that
$T_G\gtrsim {\bar T}$: if $T_G$ were not very close to ${\bar T}$,
the domains of the symmetric phase -- the sources of baryon-number excess, would
have plenty of time to shrink into oblivion before ${\bar T}$ could be reached.

Clearly, we require an estimate of $T_G$ and $\gamma_+(T_G)$.
To proceed, we assume that the fluctuations are spherical, and consider
fluctuations of amplitude $\phizc$ ($\phipc$)
and spatial size, $R\geq \xi$, the correlation length, with profiles
\bea
\phi_{0\rightarrow c}(r) & = & \phizc e^{-r^2/R^2} \ ,
\nonumber \\
\phi_{+\rightarrow c}(r) & = & (\phipc-\phi_+)e^{-r^2/R^2}
+\phi_+~.
\label{profiles}
\eea
This {\it ansatz} has been shown to be {\it quantitatively} in agreement with
numerical simulations \cite{GHK 97}.
These fluctuations are drawn from a Gibbs distribution,
\be
G_{\stackrel{+\rightarrow 0}{0\rightarrow +}} =
A\exp\left(-\frac{F_{\stackrel{+}{0}}}{T}\right) \ ,
\label{gibbs}
\ee
where $F_{\stackrel{+}{0}}$ is the free energy of the respective fluctuation.
The prefactor $A$ is expected to be small, within the range
$1 \leq A \leq 100$, although its calculation for subcritical bubble
nucleation is still an open question.
A given fluctuation $\phi$ away from a local minimum of the free-energy
density has a free energy ``cost''
\be
F(\phi)=4\pi\int_0^{\infty}
dr\, r^2\left[\frac{1}{2}\left(\frac{d\phi}{dr}\right)^2
+V(\phi)\right] \ .
\ee
With the {\it ansatz} above, these free energies take the form
\be
F_{\stackrel{+}{0}}=\alpha_{\stackrel{+}{0}}R+\beta_{\stackrel{+}{0}}R^3 \ ,
\label{freeenergies}
\ee
where,
\bea
\alpha_0(\phi) & = & \frac{3\pi\sqrt{2\pi}}{8}\phi^2_{0c} \ ,
\label{zeroconsts} \\
\beta_0(\phi) & = & \pi^{3/2}\phi^2_{0c}\left[\frac{\sqrt{2}}{4}D(T^2-T_2^2)
-\frac{\sqrt{3}}{9}ET\phi_{0c} +\frac{\lambda}{32}\phi^2_{0c}\right] \ ,
\nonumber
\eea
\bea
\alpha_+(\phi) & = & \frac{3\pi\sqrt{2\pi}}{8}(\phipc-\phi_+)^2 \ ,
\label{plusconsts} \\
\beta_+(\phi) & = & \pi^{3/2}(\phipc -\phi_+)\left[
\frac{\sqrt{2}}{4}C_1(\phipc -\phi_+) \right. \nonumber \\
& + & \left.
\frac{\sqrt{3}}{9}C_2(\phipc -\phi_+)^2 +
\frac{\lambda}{32}(\phipc -\phi_+)^3\right] \ , \nonumber
\eea
with $C_1\equiv [D(T^2-T_2^2)-3ET\phi_+
+(3/2)\lambda\phi^2]$, and $C_2\equiv (-ET+\lambda\phi_+)$.

The correlation length in the broken phase,
$\xi^2_+ \equiv [V''(\phi_+)]^{-1}$, is
\be
\frac{1}{\xi^2_+}=
4D(T^2-T_2^2)\left[\zeta^2\left(1+\sqrt{1-\frac{1}{\zeta^2}}
\right)-1\right] \ ,
\label{corrlength}
\ee
where $\zeta^2\equiv 9E^2T^2/[8\lambda D(T^2-T_2^2)] >1$.
Thus, with (\ref{zeroconsts}), (\ref{plusconsts}), and (\ref{corrlength}), we
may calculate $T_G$ via (\ref{ginzburg}).

What remains is to compute $\gamma_+(T_G)$.
We proceed by considering the Boltzmann equation for the production
of bubbles of the two competing phases, a procedure valid below percolation.
This is given by \cite{GHK 97}:
\be
\frac{\pd f_+^d}{\pd t} + 3\frac{{\dot a}}{a}f_+^d =
-|v|\frac{\pd f_+^d}{\pd R} + (1-\gamma_+)G_{0\rightarrow +} -
\gamma_+ G_{+\rightarrow 0} \ ,
\ee
where $v$ is the shrinking velocity of the subcritical bubbles, $a(t)$ is the
cosmic expansion factor, and
the volume-fraction in the broken phase is given by
\be
\gamma_+ (t) \simeq \int_{\phi_{\rm max}}^{\infty} \int_{R_{\rm min}}^{\infty}
\left(\frac{4\pi R^2}{3}\right) f_+^d(R,\phi,t)\, d\phi dR \ ,
\ee
where $f_+^d\equiv \partial n_+/\partial R\partial \phi$ is the distribution
function of disconnected domains (hence the $d$) of the broken phase modeled
by the subcritical bubbles, and $n_+$ is their number density. This equation
can be integrated by defining the quantity $Y_+\equiv f_+^d/s$, where
$s={{2\pi^2}\over {45}}g_*T^3, $ is the entropy density, and looking for
equilibrium solutions $\dot Y_+=0$.
The solution can be written as
\be
\gamma_+=\frac{I_1(\phi_{\rm max},R_{\rm min})}{1+
I_1(\phi_{\rm max},R_{\rm min})
+I_2(\phi_{\rm max},R_{\rm min})} \ ,
\ee
where
\bea
I_{\stackrel{1}{2}}(\phi_{\rm max},& R_{\rm min} &) \equiv \frac{1}{|v|}
\int_{\phi_{\rm max}}^{\infty} \int_{R_{\rm min}}^{\infty}
\int_R^{\infty} d\phi dR dR' \nonumber \\
& & \left(\frac{4\pi}{3}\right)R^3
G_{\stackrel{0\rightarrow +}{+\rightarrow 0}}(R',\phi) \ .
\eea

Using (\ref{profiles})-(\ref{freeenergies}), we get an expression for
the volume fraction $\gamma_+(T_G)$. An analytical solution can be
obtained if we neglect the cubic term in
(\ref{freeenergies}), which is a good approximation for small enough bubbles,
far from percolation.
The remaining integrals are then Gaussian, and can be performed to give
\bea
I_1&(&\phim,R_{\rm min},R_{\rm max}) = \frac{4\pi}{9|v|}A\phim\Tm^5\left \{
\left [2+2\rho \right . \right .
\nonumber \\
& & \left . \left . +\rho^2
+\frac{1}{3}\rho^3-
\frac{2}{3}\rho^4\right ]e^{-\rho}-
\rho^{9/2}{\rm erfc}\left(\sqrt{\rho}\right)
\right \}^{R_{\rm min}}_{R_{\rm max}} \ ,
\eea
where, $\rho\equiv \frac{R}{\Tm}$ and $\Tm\equiv8T/(3\pi\sqrt{2\pi}\phim^2)$
is the thermal length-scale for the system which emerges from the calculation.
An identical expression holds for $I_2$, but with $\phim$ replaced by
$\Delta_+\equiv |\phim-\phi_+|$, and $\Tm$ replaced by
$\Tm^+\equiv 8T/(3\pi\sqrt{2\pi}\Delta_+^2)$. The ratio $A/|v|$ can be obtained
numerically, as in \cite{GHK 97} or, in principle,
analytically, although we leave it as a free parameter here.
\begin{figure}[tbp]
\caption{\label{gammasm}
$\gamma_+(T_G)$ for the standard model.
}
\epsfxsize = 0.85 \hsize \epsfbox{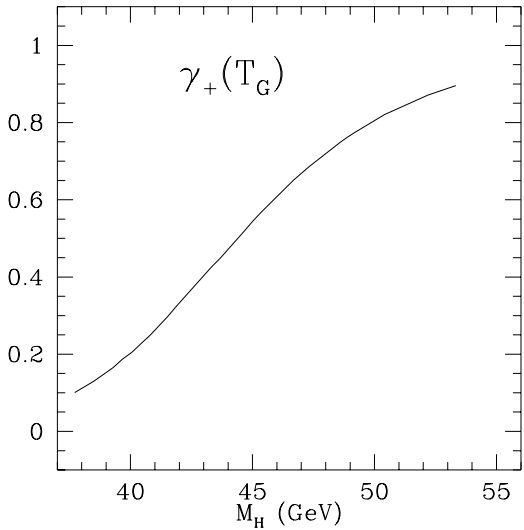}
\end{figure}

\begin{figure}[tbp]
\caption{\label{gammamssm}
$\gamma_+(T_G)$ for the MSSM. The two curves
are for $A/|v| =400$ (continuous line) and $1000$ (dotted line).
}
\epsfxsize = 0.85 \hsize \epsfbox{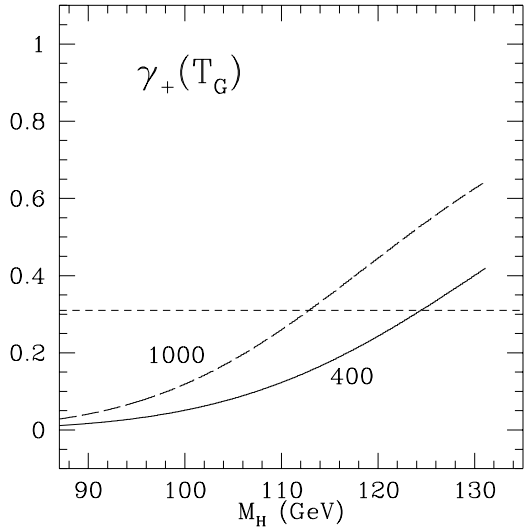}
\end{figure}

Using these expressions, we can estimate $\gamma_+(T_G)$ for the SM
and for the MSSM. We plot the results in figures
(\ref{gammasm}) and (\ref{gammamssm}), respectively.

We can now refer back to figures (1)-(4) and analyze the different
scenarios. There are two different issues that should be
discussed: first, the mass range for which the condition
$\langle\phi(T_G)\rangle/T_G\geq 1$ is satisfied; and second, how
$\gamma(T_G)$ compares with the percolation threshold probability,
$p_{\rm perc} = 0.31$. For the MSSM, the case of most immediate
interest, $\langle\phi(T_G) \rangle/T_G \geq 1$ for $m_H^{\rm
MSSM} \leq 116$ GeV (Figure 2), while $\gamma(T_G) = 0.31$ at
$m_H^{\rm MSSM}= 124$ GeV, for $A/|v|=400$ (Figure 4). Thus, in
this scenario, the broken-symmetric phase does not percolate
within the range of Higgs masses for which
$\langle\phi(T_G)\rangle/T_G \geq 1$. The allowed mass range for
baryogenesis is extended to at least $m_H^{\rm MSSM}=116$ GeV. This is
a lower bound because, as
we remarked earlier, even if $\langle\phi(T_G)\rangle/T_G < 1$, it
is possible for the condition $\langle\phi(T)\rangle/T \geq 1$ to
be achieved at a lower temperature ${\bar T}$, as long as ${\bar
T}$ is fairly close to $T_G$ (otherwise the transition would be
completed before ${\bar T}$ and no net baryon excess would be
produced). Thus, it is possible, in principle, to extend the
allowed mass range even further than the value dictated by the
condition $\langle\phi(T_G)\rangle/T_G > 1$, although a detailed
analysis of the dynamics of the phase transition is required in
order to make a more quantitative statement. An analogous
calculation for the SM yields an allowed mass range extended to at
least $m_H^{\rm SM}=45$ GeV. We should comment on the choices for
the values of $A/|v|$. The values for the prefactor of the
subcritical bubble nucleation rate ($A$) or its shrinking velocity
in a plasma ($v$) are not known. However, from results for
critical bubble nucleation and expansion in the electroweak
transition, it is reasonable to expect that $1 \leq A \leq 100$
and $0.005 < v < 0.5.$ We are treating $A/|v|$ as a free parameter
(fortunately, only the ratio of the two is relevant), and values
between $100$ and $1000$ seem to be reasonable, judging from the
ranges of $A$ and $|v|$. Hopefully, detailed simulations of weakly
first order transitions could further restrict the range of
$A/|v|$.

We have described how electroweak baryogenesis can proceed even if the
phase transition is weakly first order. This idea is radically different to
existing lore about such scenarios. In weak phase transitions, baryon number
violating processes are sufficiently copious in the broken phase that any
baryon excess is washed out at $T_c$. However, the important point is that the
transition dynamics can be such that, at a lower temperature,
when baryon number violation becomes exponentially suppressed in the broken
phase (at ${\bar T}$), the transition may not yet have
completed.
There then remains the
possibility for a sufficient baryon excess to be produced during the
remainder of the transition.
Thus, this approach may have important ramifications for electroweak
baryogenesis.
In this letter we have described how to estimate
the relevant quantities in these scenarios analytically.
In a later publication we will present a detailed semi-analytical and
numerical approach in which we will obtain more precise values. This will
involve taking into account the full form of the free energy, the effects of
nonabelian gauge interactions, and an accurate
analysis of the phase transition dynamics and freeze-out rates.

We thank Antonio Riotto for extensive discussions. M.G. was partially
supported by the National Science Foundation, grant no. PHYS-9453431.
He thanks the High Energy
Group at Boston University, where part of this work was completed, for their
hospitality. M.T. was
supported by the U.S. Department of Energy.

\end{document}